\documentclass[namedreferences]{solphy_emu}
\usepackage[optionalrh]{spr-sola-addons_emu} 
\usepackage{graphicx}                    
\usepackage{color}                       
\usepackage{url}                         

\newcommand{\aap}{    {\it Astron. Astrophys.}}

\newcommand{\apj}{    {\it Astrophys. J.}}

\newcommand{\solphys}{{\it Solar Phys.}}

\begin{document}

\begin{article}

\begin{opening}

\title{Subsurface Meridional Circulation in the Active Belts}
%
\author{I.~\surname{Gonz{\'a}lez Hern{\'a}ndez}$^{1}$\sep
       S.~\surname{Kholikov}$^{1}$\sep
       F.~\surname{Hill}$^{1}$\sep  
       R.~\surname{Howe}$^{1}$\sep  
       R.~\surname{Komm}$^{1}$\sep  
       }

%
\runningauthor{I. Gonz{\'a}lez Hern{\'a}ndez {\it et al.}}
\runningtitle{Subsurface Meridional Circulation in the Active Belts}

%
\institute{$^{1}$ National Solar Observatory, Tucson, AZ, USA;
                     email: \url{irenegh@nso.edu} \\ 
             }

\begin{abstract}
Temporal variations of the subsurface meridional flow with the solar cycle have been reported by several authors. The measurements are typically averaged over periods of time during which surface magnetic activity existed in the regions were the velocities are calculated. The present work examines the possible contamination of these measurements due to the extra velocity fields associated with active regions plus the uncertainties in the data obtained where strong magnetic fields are present. We perform a systematic analysis of more than five years of GONG data and compare meridional flows obtained by ring-diagram analysis  before and after removing the areas of strong magnetic field. The overall trend of increased amplitude of the meridional flow towards solar minimum remains after removal of large areas associated with surface activity. We also find residual circulation toward the active belts that persist even after the removal of the surface magnetic activity, suggesting the existence of a global pattern or longitudinally-located organized flows.

\end{abstract}

%

\end{opening}
%
\section{Introduction}

Meridional circulation is a major focus of solar studies as it has become a  key component of solar-dynamo models \cite{miesch2005}. Its temporal variation has been shown to influence the upcoming solar-cycle features \cite{dikpati2004}. Thus, accurately describing the meridional flow both in latitude and depth, as well as understanding its variation with the solar cycle has become a focus of local-helioseismology studies.

\begin{figure} 
\centerline{\includegraphics[width=0.97\textwidth,clip=]{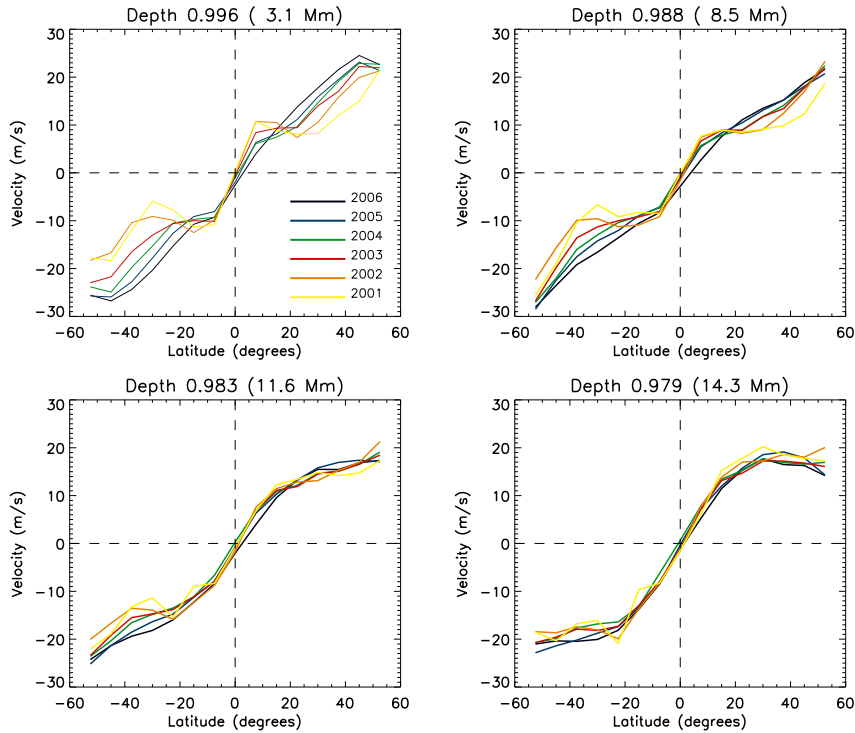}}
\caption{Yearly averages of the meridional flow obtained by ring-diagram analysis of GONG continuous set of data at four different depths. The variation with the solar cycle clearly observed at the superficial layers is less pronounced at deeper layers.}\label{fig:meridional}
\end{figure}

Early studies using magnetic-feature tracking \cite{komm1993,snodgrass1996} and surface Doppler measurement \cite{hathaway1996, nesme1997} found evidence of poleward meridional flow at the solar surface and variations of the flow with the solar cycle. Helioseismic observations, which are able to infer the depth profile of the flows, confirmed the existence of such a flow \cite{giles1997}. An average poleward flow at the surface and upper layers of the convection zone of 10\,--\,20\,m\,s$^{-1}$ seems to be the consensus.
\inlinecite{giles2000} was the first to use a local helioseismology technique, time-distance \cite{duvall1993} to study the temporal evolution of the meridional flow. The results showed a variation of the amplitude of the flow which decreased towards solar maximum during cycle 23. \inlinecite{chou2001} extended the results also finding an increase on the amplitude towards solar minimum using data obtained during the previous cycle. Ring-diagram analysis of Michaelson Doppler Imager (MDI) data \cite{haber2002, basu2003}, during the rising phase of cycle 23, and Global Oscillation Network (GONG) data \cite{gonzalez2006, zaatri2006}, in the declining phase of the same cycle, confirmed the overall  variation of the meridional flow with the solar cycle, with larger amplitude toward solar minimum. \inlinecite{ulrich2005} show temporal variations of the meridional flow for almost two solar cycles using a technique that tracks points at the surface. They also found an anticorrelation between the amplitude of the meridional flow and the solar cycle.
 
The  particular behavior of the meridional flow around the latitude of magnetic activity concentration has been noted by most of these authors. A steep gradient of the flow around the Equator toward the activity latitudes,  which increases with increasing surface magnetic activity,  has been a common factor in these observations. 

\inlinecite{gizon2003} analyzed two Carrington rotations eliminating the contribution from areas of high magnetic activity and found a marked difference from the meridional flow calculated using only quiet areas in the activity latitudes.
This result suggested that the organized inflows that had been shown to exist surrounding large active region complexes \cite{zhao2001,haber2004,komm2004,braun2004} were responsible for this observed component of the meridional flow that converges towards the active latitudes. \inlinecite{zhao2004} pointed out that there may be an extra component in the active belts beyond the inflows surrounding active complexes, since they found residual flows below the maximum depth of the flows associated to the activity.

Six years of high-resolution observations from GONG (\url{http://gong.nso.edu/data}),  during the declining phase of solar cycle 23, give us the unprecedented opportunity to study the meridional flow using local helioseismology with continuity. It also allows us to study the effect of the surface activity on the inferred flow with statistical significance. Figure \ref{fig:meridional} shows the yearly averaged meridional flow obtained by applying the standard ring-diagram analysis \cite{haber2002} to the continuous set of GONG high-resolution data. The bumps or inflows around the active belts are clearly visible in the figures. In this work we discriminate between measurements of meridional flow averaging all available data and those obtained by using only areas of quiet Sun to investigate how the surface magnetic activity affects the inferred flows.

\section{Data Analysis}

We apply standard ring-diagram analysis to GONG high-resolution Dopplergrams from July 2001 to December 2006 to infer the meridional flow from the solar surface to a depth of approximately 16\,Mm for the declining phase of cycle 23.

The ring-diagrams method studies high-degree waves propagating in localized areas over the solar surface to obtain an averaged horizontal-velocity vector for that particular region. By analyzing a mosaic of these patches, it is possible to develop a three-dimensional velocity map in the depth range where the waves propagate. Typical ring-diagram analysis uses 1664-minute series of full-disk Dopplergrams  with a resolution of about 1.5 Mm per pixel at the center 
of the disk. Patches of 16$^{\circ}$ square, apodized to 15$^{\circ}$--diameter areas, are tracked at the surface rotation rate. A three-dimensional FFT is applied to a tracked area, and the corresponding power 
spectrum is fitted using a Lorentzial profile model that includes a frequency shift term due to the horizontal-velocity flow (\opencite{haber2002}). Finally, the fitted velocities are inverted using a least-squares method to recover the depth dependence of the velocity flows. The GONG ring-diagram pipeline has been used for the work presented here. Details of the pipeline can be found in \inlinecite{corbard2003}. A single horizontal-velocity vector ($v_{x},v_{y}$) at several depths is obtained as the result of analyzing a single patch.

To study the meridional circulation, we concentrate on the $v_{y}$ component of the calculated flows. Local-helioseismology inferred flows at high latitudes have been shown to be affected by the periodic variation of the solar $B_{0}$-angle \cite{gonzalez2006, zaatri2006}. We do not yet have a full understanding of the effect. \inlinecite{beckers2007} analyzed the problem and showed that a different correction can result in substantially different results. Hence, we have decided to not correct the inferred flows. Since the $B_{0}$ angle effect will affect equally the yearly averages of all  calculated flows, the temporal variations that we are interested in should still be valid.

Thanks to the continuous stream of data provided by the GONG instrument, we have obtained velocity flows for 5.5  uninterrupted years. The yearly-averaged meridional flow at four particular depths can be seen in Figure~\ref{fig:meridional}. Close to the surface, at approximately 3\,Mm depth, the marked increase in the amplitude of the flows as the solar cycle progresses towards minimum can be seen, confirming previous results. The bumps around the active belts are very pronounced at this depth, and the amplitude decreases with decreasing magnetic activity.
With the limited resolution attained by using a standard ring-diagram analysis, the variation of the flows at high latitudes could also be interpreted as an effect due to the contribution of organized flows around active regions.

\begin{figure} 
\centerline{\includegraphics[width=0.98\textwidth,clip=]{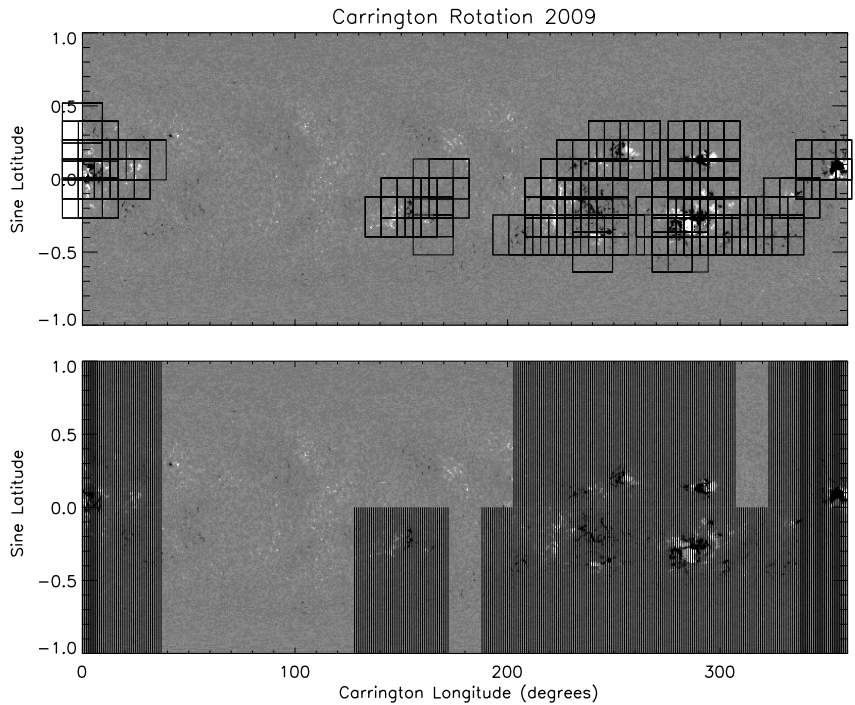}}
\caption{Two different approaches to eliminate data associated with surface magnetic activity. This figure shows the two different type of masks applied to Carrington rotation 2009. Top panel shows mask 1: only those patches associated with surface magnetic activity are eliminated. Bottom panel shows mask 2: all patches in the same hemisphere and at the same longitude as a particular patch with surface magnetic activity are eliminated. }\label{fig:removingactivity}
\end{figure}

With the spatial resolution of the standard ring-diagram analysis, removing areas of surface magnetic activity to measure meridional circulation in the quiet Sun is complicated, especially in periods of high activity. However the period from 2003 to 2004 of the complete data set presents areas of high activity combined with large areas of very low activity, allowing for statistically significant comparison between meridional flows averaged over all areas versus those obtained only from quiet regions.

\begin{figure} 
\centerline{\includegraphics[width=0.97\textwidth,clip=]{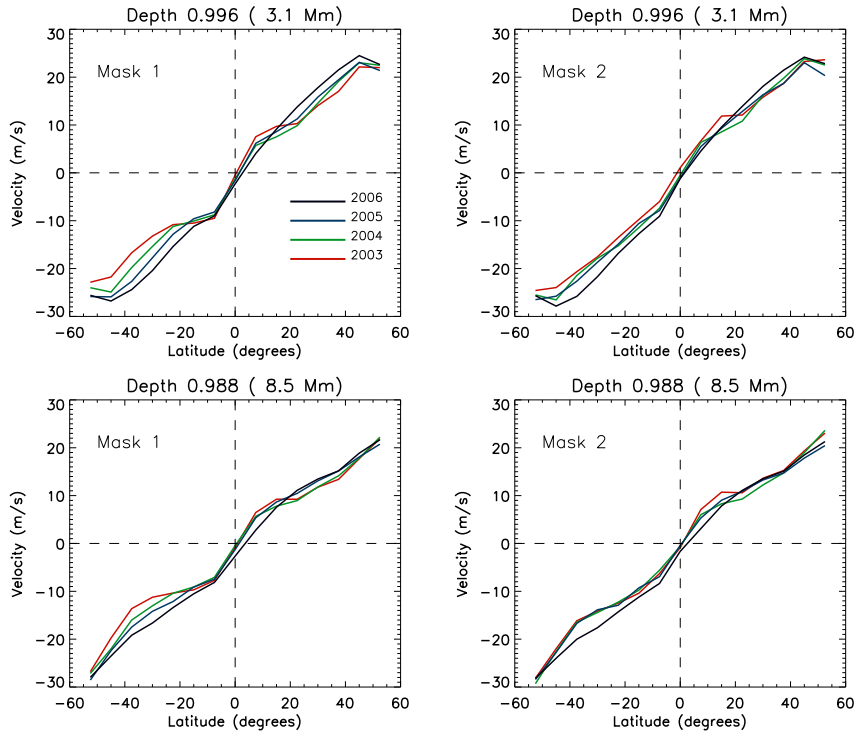}}
\caption{Yearly averages of meridional flows obtained by ring-diagram analysis of GONG continuous set of data at two different depths after applying mask 1 (left) and mask 2 (right). The variation with the solar cycle is attenuated after applying the more aggressive masking procedure.}\label{fig:aftermer}
\end{figure}

\begin{figure} 
\centerline{\includegraphics[width=0.9\textwidth,clip=]{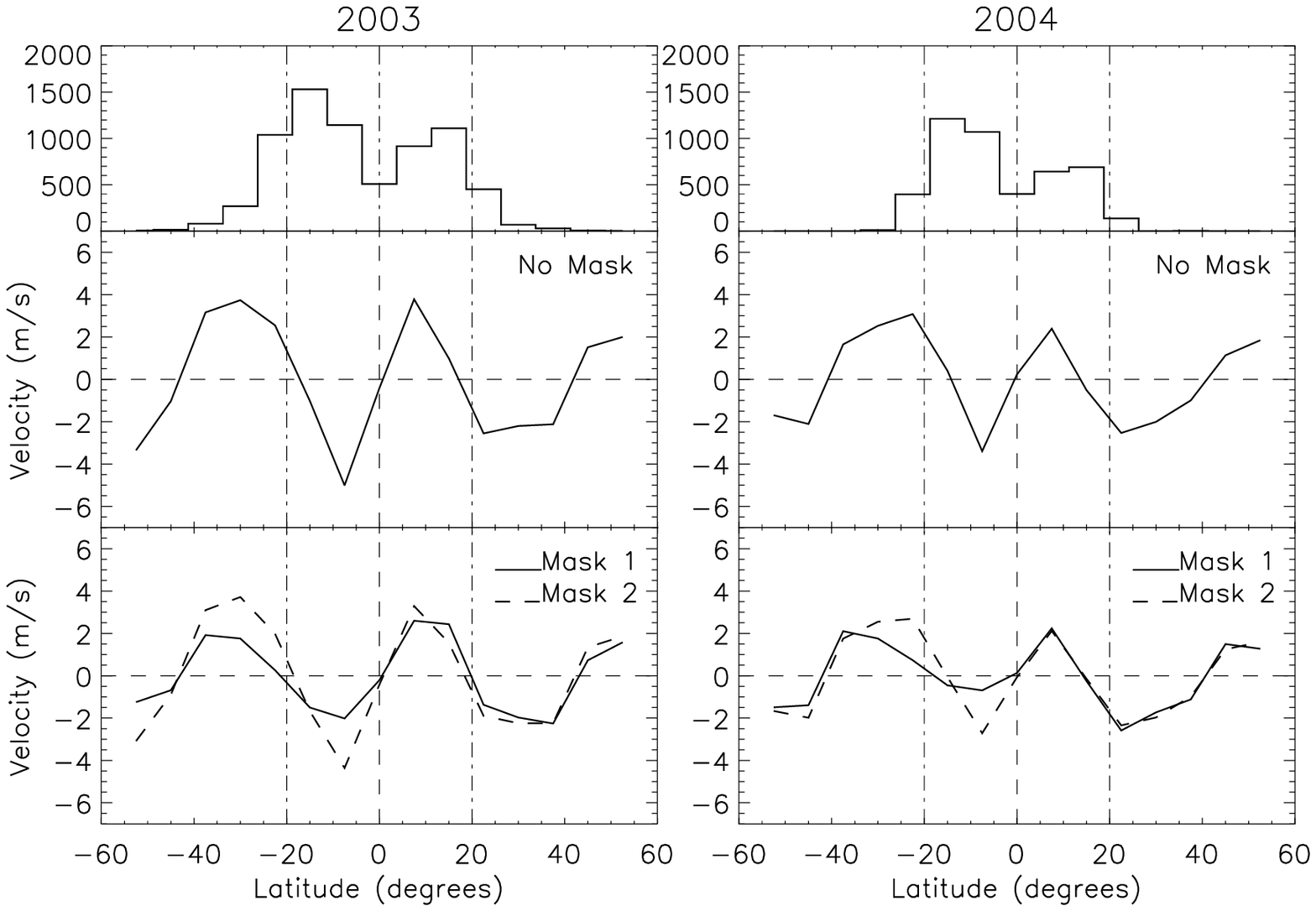}}
\caption{Residuals from meridional circulation around the active belts for years 2003 and 2004 at a depth of 5.8\,Mm. Top panels show the number of ring-diagram patches eliminated because of surface magnetic activity. Middle panels show the residuals when the meridional flow is calculated using all available data. Bottom panel show residuals after eliminating patches associated to surface activity using mask 1 (discontinuous line) and mask 2 (solid line)}\label{fig:bumps1}
\end{figure}

\begin{figure} 
\centerline{\includegraphics[width=0.9\textwidth,clip=]{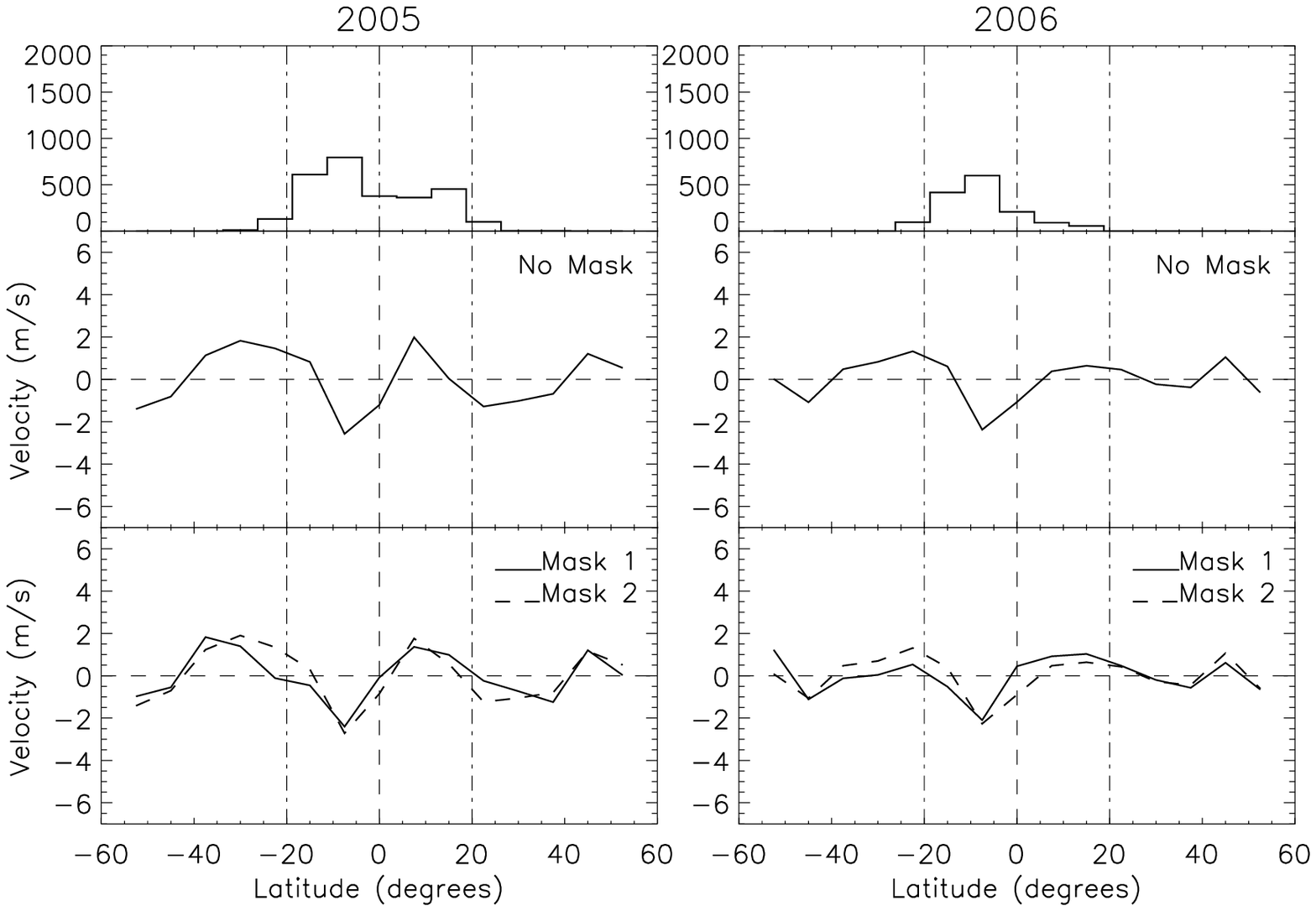}}
\caption{Same as Fig~\ref{fig:bumps1} for years 2005 and 2006}\label{fig:bumps2}
\end{figure}

Our first attempt to isolate quiet areas from active ones was based on removing all of the patches with an averaged magnetic-field strength above a certain threshold (mask 1). For our purposes, 10\,Gauss seemed to account for most of the surface activity. The top panel of Figure~\ref{fig:removingactivity} shows the areas removed using this approach. The square patches corresponds to the ring-diagram standard areas. After removing these patches, the residual meridional flow was practically the same as that obtained by including all areas (Figures~\ref{fig:aftermer} - ~\ref{fig:bumps2}).

The second attempt  removed data associated with surface activity more aggresively. It consists of removing the patches with average-magnetic field strength above 10\,Gauss and all the patches at the same longitude and in the same hemisphere as the masked one (mask 2)(see bottom panel of Figure~\ref{fig:removingactivity}). In this way, flows associated with surface activity that extend for long distances will be completely removed. We made the assumption here that organized flows around a particular active region do not cross the Equator. This is not completely accurate when the activity is very near the Equator, but statistically we expect the flows crossing the Equator in both directions to cancel out. The removed areas are dark colored in the bottom panel of Figure~\ref{fig:removingactivity} for Carrington rotation 2009. In both cases, the averaged magnetic field for a particular patch has been calculated using MDI magnetograms.

The meridional flow obtained for years 2003 to 2006 at two particular depths before and after removing the surface areas of activity is presented in Figure~\ref{fig:aftermer}. The overall trend of the flow, increasing as the activity decreases does not change. However, the trend is less prominent after applying mask 2.

To focus on the active-belt extra circulation, Figures~\ref{fig:bumps1} and ~\ref{fig:bumps2} shows the residual meridional flow after removing a second-order polynomial from the inferred flows before (center) and after (bottom) removing activity using  mask 1 (dashed line) and mask 2 (solid line) for a depth of 5.8\,Mm. The reason we remove a low-order polynomial instead of a temporal average of the results is to avoid the dominance of the extra circulation bumps in the average. Using the average of the year with minimum activity (in this case 2006) was also considered, but that year shows some asymmetry around the Equator and the results would be difficult to interpret.
The histograms at the top of the figures represent the number of ring-diagram patches removed at different latitudes. In a way, they are a proxy of the surface activity at our resolution. The southern hemisphere has been more active during the second half of the cycle, which can be easily seen in the figures.
The figure shows that the inflows are almost unaffected by applying mask 1 but are changed when applying  mask 2. However, not all of the extra circulation goes away when applying even the most aggressive masking procedure, especially at depths around 6.0\,Mm.

\begin{figure} 
\centerline{\includegraphics[width=0.9\textwidth,clip=]{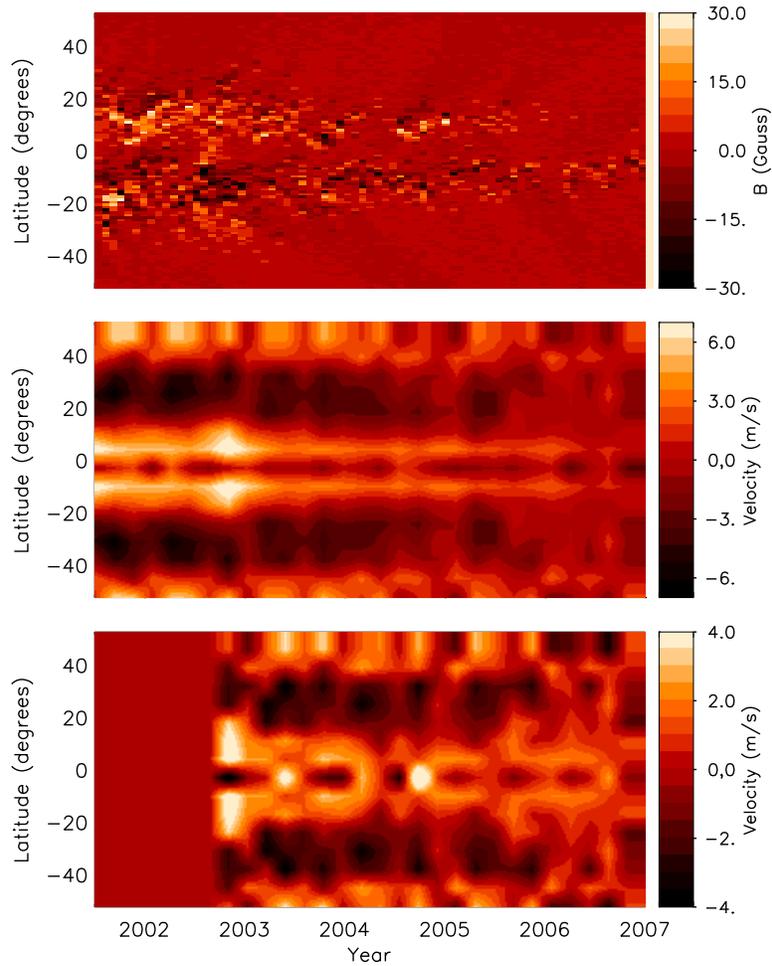}}
\caption{Temporal variation of the meridional circulation residuals when averaging all available data (center) and after applying mask 2 (bottom). Both figures show the flow at a depth of 5.8\,Mm. averaged over the two hemispheres. Positive velocities are directed towards the center of activity. The top panel shows the magnetic activity during the same period (calculated from MDI synoptic magnetograms).}\label{fig:butterfly}
\end{figure}

The middle panel of Figure~\ref{fig:butterfly} shows  the temporal variation  of the residuals after removing a low-order polynomial to the inferred meridional flow, when all the available data is considered. The results of the north and south hemisphere have been averaged, so the image is symmetric.
The bottom panel shows the residuals after applying   mask 2. It is clear that a second-order residual remains even after the removal of most surface activity.


\section{Discussion}

We find a strong variation of the meridional flow with the solar cycle in the shallow layers of the convection zone (Figure~\ref{fig:meridional}). Previous authors have reported a deceleration of the flow toward solar maximum \cite{giles2000,haber2002,basu2003,zhao2004} when examining data obtained during the rising phase of solar cycle 23. We complete the cycle using continuous observations from GONG and find that the flows accelerate during the second half of the cycle confirming the results that \inlinecite{chou2001} obtained for the late years of cycle 22. When averaging all of the available data, including areas of surface magnetic activity, the temporal variation depends on latitude and presents a maximum difference of $\approx$\,10\,m\,s$^{-1}$ around 40$^{\circ}$ between solar minimum and solar maximum. Towards the interior of the Sun, the variation is less pronounced, with a maximum difference of approximately 5\,m\,s$^{-1}$ at a depth of 10\,Mm. The variation is greater in the southern hemisphere, coinciding with the surface activity. Even after removing all of the surface activity (mask 2), the variation of the flow towards solar minimum persists, but the acceleration is smaller.

The meridional flow variation with the solar cycle is less pronounced at deeper depths. At 14\,Mm the difference between the yearly averaged flows is still visible at high latitudes (see Figure~\ref{fig:meridional}), but the trend is no longer obvious. \inlinecite{gizon2008} show independent observations of the meridional flow at the surface and at a depth of 60\,Mm and find that the time varying component of the flows at the deeper layer has the opposite sign to that at the surface. They also present a model that qualitatively accounts for the variation of the meridional flow at these two depths and the fact that they are anticorrelated. Although our analysis only returns information over the upper 15\,Mm, the depth dependence of the results is such that it would be consistent with their findings, with a reversal in the time varying component at depths below our accessible range.

We confirm the extra circulation in the active belts previously reported by other authors. \inlinecite{spruit2003} presented a model that explained the torsional oscillation as a geostrophic flow due to the lower subsurface temperature in active regions. This model predicts the appearance of flows from the edges towards the center of the main latitude of magnetic activity, a meridional version of the torsional oscillation with a maximum amplitude of $\approx$\,6\,m\,s$^{-1}$ at the surface. The model also predicts a rapid decline of these oscillations with depth, which would disappear below 30\,Mm. However, our results show a more rapid attenuation of the inflows in the active belt, which disappear around 10--14\,Mm.

The observed extra circulation in the active latitudes varies with the solar cycle, decreasing towards solar minimum. These inflows towards the center of activity do not disappear when the contribution of the surface activity is removed (see Figures~\ref{fig:bumps1} and ~\ref{fig:bumps2}), although the amplitude is  reduced  when only data from the quiet Sun is used for the calculation of the flows. In a preliminary study of two single Carrington rotations \cite{gonzalez2007} the meridional flow obtained  in the active latitudes when using only the quiet Sun differs from that calculated using all available data. This confirmed the results from \inlinecite{gizon2003}. These limited studies suggested that the extra circulation was due to the inflows that have been shown to exist surrounding large active complexes \cite{zhao2001,haber2004,komm2004,braun2004}. However, the systematic study that we present here shows that this extra circulation cannot be accounted for exclusively by localized inflows associated with surface activity, since it persists even when using only data from the quiet Sun. 

\inlinecite{chou2005} applied  the time-distance technique  to Taiwan Oscillation Network (TON) data from 2000 to 2003. They studied the temporal variation of the meridional circulation down to a depth of 0.793\,$R_{\odot}$, focusing on the divergent flow around the activity. They concluded that the amplitude of the divergent flow increases with depth in the upper layers of the convection zone, with a maximum amplitude at layers shallower than 0.9\,R$_{\odot}$, reversing the trend underneath. This is compatible with our results. They also found that for greater depths, the inflows in the active belts are such that the meridional flow changes signs, allowing for countercirculation.

\inlinecite{svanda2007} calculate the meridional flow transport speed from synoptic maps of the longitudinal magnetic field and compare it with results from time-distance helioseismology for particular Carrington rotations. The main conclusion of the paper is that the meridional flow derived from helioseismology is affected by localized activity, as compared to the meridional flux transport and masking surface activity renders closer results. We show here that the masking of activity changes the flow in the active belts, but a second component to the meridional circulation remains even after the exclusion of large amounts of data associated to surface activity. In fact, the results from flux transport showed by \inlinecite{svanda2007} (Figure 2) agree very well with our results, showing an overimposed circulation in the active latitudes for the Carrington rotations studied during high activity with local maximum below 20$^{\circ}$ and local minimum above 30$^{\circ}$ (see Figure~\ref{fig:bumps1}). In a recent paper, the same authors \cite{svanda2008} concluded that the mass flows around active regions significantly affect the meridional flows. Our results agree with that conclusion; hence we show here the mean meridional flow for the declining phase of solar cycle 23 after removing the contribution from surface activity. They also found that local flows are more important in areas of leading magnetic polarity. With ring-diagram analysis, we do not have enough resolution to resolve the active region, so we have taken the approach to remove as much of the area sourrounding active regions as possible to eliminate the contribution from local flows.

\section{Conclusions and Future work}

We find an acceleration of the meridional flow in the upper layers of the convection zone in the second half of cycle 23. Our results confirm the variation  of the meridional flow with the solar cycle, having the amplitude decreasing as the magnetic activity increases. This variation persists after removal of the data associated with surface activity, although is less pronounced. Below 10\,Mm the temporal variation is only  clearly seen above 30$^{\circ}$. In the depth range surveyed by our analysis (0\,--\,15\,Mm) we do not observe a reversal of the sign of variation, however the attenuation of the temporal variation with depth is consistent with a reversal at deeper depths as suggested by \cite{gizon2008}

The superimposed circulation  in the active belts persists after removal of the surface activity and large areas surrounding the activity. However the amplitude of the inflows is attenuated. This implies that organized flows around large active regions are not exclusively responsible for the extra circulation, although they modulate it by moving the average towards the average latitude of activity. The meridional flow in shallow layers is more affected by surface activity than at deeper layers, which implies that the inflows associated to the activity are confined to the upper layers of the convection zone.

When using only quiet-Sun areas to calculate the meridional flow, the mean latitudinal position of the extra circulation in the active belts does not change significantly during the studied period of 2001\,--\,2006. Previous work, using data from 1996 to 2001, seems to find a migration of the features that follows the mean location of activity. An explanation of this may be that the activity migrates more slowly towards the Equator in the second half of the solar cycle. A similar effect is seen in the torsional oscillation of the zonal flow, which has a rapid migration towards the Equator at the beginning of the cycle but slows down in the second half \cite{howe2006}.

There are several explanations for the second component of the meridional circulation that persist even after aggressively masking surface activity. First, the inflows at the active latitudes are a global pattern in the active belts. Second, the organized flows around activity are also important for much weaker magnetic areas. Third, the organized flows persist well after a strong active region have become diffuse. Finally, the concept of active longitudes proposed long ago by \inlinecite{becker1955} and more recently revisited by \inlinecite{detoma2000} may also be an explanation, if the flow associated with active regions in the active longitudes persist for consecutive episodes of active region formation. With the limited resolution of the standard ring-diagram analysis, the discrimination between the possible explanations is a challenge, due to the lack of enough data for a statistically significant analysis. Other local helioseismology techniques, such as  high resolution ring-diagrams \cite{hindman2004}, time-distance \cite{zhao2004b} or seismic holography \cite{braun2004},  although computationally more expensive, provide higher resolution which would help to see the full picture.

%
\begin{acks}
The authors thank L. Gizon, C. Lindsey and A. Zaatri for helpful discussions.
This work utilized data obtained by the Global Oscillation Network Group (GONG) program, managed by the National Solar Observatory, which is operated by AURA, Inc. under a cooperative agreement with the National Science Foundation. The data were acquired by instruments operated by the Big Bear Solar Observatory, High Altitude Observatory, Learmonth Solar Observatory, Udaipur Solar Observatory, Instituto de Astrof{\'{\i}}sica de Canarias, and Cerro Tololo Interamerican Observatory. Magnetograms from the SOI/MDI on SOHO have also been used for the analysis. SOHO is a project of international collaboration between ESA and NASA.
\end{acks}

%
%
%

\end{article} 
\end{document}